\documentclass[12pt]{article}
\usepackage{amsmath,amssymb}
\usepackage{epsfig} 
\usepackage{cite}

\def\m@th{\mathsurround=0pt }
\def\leftrightarrowfill{$\m@th \mathord\leftarrow \mkern-6mu
	\cleaders\hbox{$\mkern-2mu \mathord- \mkern-2mu$}\hfill
	\mkern-6mu \mathord\rightarrow$}

\def\overleftrightarrow#1{\vbox{\ialign{##\crcr
	\leftrightarrowfill\crcr\noalign{\kern-1pt\nointerlineskip}
	$\hfil\displaystyle{#1}\hfil$\crcr}}}

\newcommand{\be}{\begin{equation}}
\newcommand{\ee}{\end{equation}}

\newcommand{\Tr}{\mathop{\rm Tr}}
\def\I{\rm 1\kern-.24em l}  

\def\shat{\ifmmode \hat{s}\else $\hat{s}$\fi}

\def\nn{\nonumber}
\def\UV{{\textsc{uv}}}
\def\IR{{\textsc{ir}}}

\newcommand{\newc}{\newcommand}

\newc{\gsim}{\lower.7ex\hbox{$\;\stackrel{\textstyle>}{\sim}\;$}}
\newc{\lsim}{\lower.7ex\hbox{$\;\stackrel{\textstyle<}{\sim}\;$}}
\newc{\ie}{{\it i.e.}}
\newc{\etal}{{\it et al.}}
\newc{\mev}{\hbox{\rm\,MeV}}
\newc{\gev}{\hbox{\rm\,GeV}}
\newc{\tev}{\hbox{\rm\,TeV}}
\newc{\xpb}{\hbox{\rm\, pb}}
\newc{\xfb}{\hbox{\rm\, fb}}

\newc{\G}{{\cal G}}
\newc{\h}{{\cal H}}
\newc{\D}{{\cal D}}
\newc{\E}{{\cal E}}

\newc{\mtop}{m_t}
\newc{\mbot}{m_b}
\newc{\mz}{M_Z}
\newc{\mw}{M_W}
\newc{\alphasmz}{\alpha_s(M_Z)}
\newc{\swsq}{\sin^2\theta_W}
\newc{\cwsq}{\cos^2\theta_W}
\newc{\tw}{\tan\theta_W}
\newc{\cw}{\cos\theta_W}
\newc{\sw}{\sin\theta_W}
\newc{\BR}{\hbox{\rm BR}}
\newc{\zbb}{Z\to b\bar}
\newc{\Gb}{\Gamma (Z\to b\bar b)}
\newc{\Gh}{\Gamma (Z\to \hbox{\rm hadrons})}
\newc{\sgn}{\mbox{sgn}}

\newcounter{mysubequation}[equation]

\def\beq{\begin{equation}}
\def\eeq{\end{equation}}
\def\bea{\begin{eqnarray}}
\def\eea{\end{eqnarray}}
\def\ZZ{\mathbb Z}
\def\slashchar#1{\setbox0=\hbox{$#1$}           
   \dimen0=\wd0                                 
   \setbox1=\hbox{/} \dimen1=\wd1               
   \ifdim\dimen0>\dimen1                        
      \rlap{\hbox to \dimen0{\hfil/\hfil}}      
      #1                                        
   \else                                        
      \rlap{\hbox to \dimen1{\hfil$#1$\hfil}}   
      /                                         
   \fi}                                         %
\catcode`@=11
\long\def\@caption#1[#2]#3{\par\addcontentsline{\csname
  ext@#1\endcsname}{#1}{\protect\numberline{\csname
  the#1\endcsname}{\ignorespaces #2}}\begingroup
    \small
    \@parboxrestore
    \@makecaption{\csname fnum@#1\endcsname}{\ignorespaces #3}\par
  \endgroup}
\catcode`@=12

\begin{document}

\baselineskip=18pt

\setcounter{footnote}{0}
\setcounter{figure}{0}
\setcounter{table}{0}

\begin{titlepage}
\begin{flushright}
UAB-FT--636
\end{flushright}
\vspace{.3in}

\begin{center}
\vspace{1cm}

{\Large \bf  Stable skyrmions from extra dimensions}

\vspace{.8cm}

{\bf Alex  Pomarol$^{a}$ and Andrea Wulzer$^{b}$}

\vspace{.5cm}

\centerline{$^{a}${\it  IFAE, Universitat Aut\`onoma de Barcelona, 08193 Bellaterra, Barcelona}}
\centerline{$^{b}${\it Institut de Th\'eorie des Ph\'enom\`enes Physiques, EPFL,  CH--1015 Lausanne, Switzerland}}

\end{center}
\vspace{.8cm}

\begin{abstract}
\medskip
\noindent

\end{abstract}
We show that skyrmions arising from compact  five  dimensional models 
have stable  sizes.
We numerically obtain  the skyrmion configurations
and calculate their size and energy.
Although their size strongly depends  on the magnitude  of localized kinetic-terms,
their energy is quite model-independent ranging between $50-65$ times 
$F_\pi^2/m_\rho$, where $F_\pi$ is the Goldstone decay constant and $m_\rho$
the lowest Kaluza-Klein  mass.
These skyrmion configurations interpolate between  small 4D  YM instantons and 
 4D skyrmions made of Goldstones and a massive vector boson.
Contrary  to  the original 4D skyrmion and previous 5D extensions,  
these configurations   have sizes larger than the inverse of the cut-off scale and therefore
they are trustable  within our effective 5D approach. 
Such solitonic particles can have interesting phenomenological
consequences as they carry a conserved topological charge analogous to
baryon number.

\bigskip
\bigskip

\end{titlepage}


\section{Introduction}
\label{intro}

Non-linear $\sigma$-models can have topological stable configurations 
known as skyrmions \cite{Skyrme:1961vq}.  For Goldstone fields parametrizing the coset $G/H$,
skyrmions can exist always that $\pi_3(G/H)\not=0$.
Nevertheless, the existence of non-singular configurations
cannot be determined within  
the non-linear  $\sigma$-model,
since their  size strongly depends on the UV-completion   of the model
  at scales around $ 4\pi F_\pi$, where
 $F_\pi$ is the Goldstone decay constant.

Five dimensional gauge theories provide  UV-completions 
of the non-linear  $\sigma$-models.
\footnote{Strictly   speaking,
5D  theories provide only    UV-extensions of the non-linear $\sigma$-models
since they are not  well-defined theories at arbitrarily high energies.}   
Any  compact 5D gauge theory, in which     the    bulk gauge symmetry $G$  is broken down to  $H$  at one boundary and to nothing at the other, is described at low-energies by a 4D  non-linear $G/H$ $\sigma$-model with 
  $F_\pi\sim \sqrt{M_5/L}$, where 
 $M_5$ is the inverse squared of the 5D gauge coupling and 
 $L$ is the compactification scale in conformal coordinates. 
The  cut-off scale of  the 5D gauge theories is estimated to be
 $\Lambda_5\sim24\pi^3 M_5$. Therefore, provided that
 $M_5\gtrsim 1/L$,   these theories are   valid
up to   energies   much larger than the scale $4\pi F_\pi$.
Being this the case, compact five dimensional gauge theories 
allows us to  address 
the question of the stability of the skyrmion configurations and calculate their properties.
This is the subject of this paper.

We will calculate numerically the  skyrmion configuration arising from compact  extra dimensional models, concentrating  in flat and AdS spaces.
We will show that  these configurations have  stable sizes,  calculable 
within the regime of validity of our effective 5D theory.
The size, however, depends not only on  $M_5$   and   $L$,
but also on the coefficient of the lowest higher-dimensional operator of the 5D theory.
The energy of the skyrmion will also be calculated and shown to be
predicted in a narrow range.

Attemps to address the stability of the skyrmion in extra dimensional models
can be found recently in the literature \cite{Son:2003et,Hong:2007kx,Hata:2007mb,Hong:2006ta,Hill:2001bt}. Nevertheless, we consider that none of them address fully consistently
the stability of the skyrmion configuration 
(to determine the size within the higher-dimensional effective theory  approach).
Our analysis  is also the first to exactly   calculate the skyrmion configuration in 5D models
that requires the use of  numerical methods.

The interest in the existence of these topological objects
is not only  theoretical but also phenomenological.
In the recent years there has been a lot of activity in using extra dimensions in order to 
 break the electroweak symmetry.
In  these examples  the existence of topologically stable particles
can lead to   important  cosmological implications.  
Another interest in skyrmion  configuration  arises in the context of   warped 5D spaces.  
The  AdS/CFT correspondence  
tells us that skyrmions in 5D are the dual of the  "baryons" of 4D strongly coupled theories.
Therefore
 studying  these 5D solitons can be useful to learn about  the properties of 
 the baryons.

\section{Skyrmions in compact and  warped 5D spaces}
\label{}

We will be looking for skyrmion configurations  arising from
the global symmetry breaking pattern   $SU(2)_L\times SU(2)_R\rightarrow SU(2)_{V}$. 
These configurations will still be valid for any  breaking  $G\rightarrow H$  always  that $G$ and  $H$ contains respectively 
$SU(2)_L\times SU(2)_R$ and $SU(2)_V$ as a subgroup.
This includes, for example,  the case $SU(N)_L\times SU(N)_R\rightarrow SU(N)_{V}$.

The  class of 5D theories  that we will be considering   
corresponds
to  the following one.
They are $SU(2)_L\times SU(2)_R$ gauge theories  with  metric 
 $ds^2=a(z)^2\left(dx_\mu dx^\mu-dz^2\right)$, where $x^\mu$ represent the usual $4$ coordinates (4D indeces are raised and lowered with the flat Minkowski metric) and $z$, which runs in the interval $[z_{\UV},z_{\IR}]$,  denotes the extra dimension. 
 We  will take $a(z)\geq a(z_{\IR})=1$.
We will denote respectively   $L_M$ and $R_M$, where $M=(\mu,5)$,  the $SU(2)_L$ 
 and $SU(2)_R$ gauge connections parametrized by 
  $L_M=L_M^a\sigma_a/2$ and $R_M=R_M^a\sigma_a/2$ where $\sigma_a$ are the Pauli matrices. The chiral symmetry breaking is imposed on the boundary at $z=z_{\IR}$
   (IR-boundary)  by 
 requiring the following boundary  conditions:
\beq
\left(L_\mu-R_\mu\right)\left|_{z=z_{\IR}}\right.=0\, ,\;\;\;\;\;\;\left(L_{\mu 5}+R_{\mu 5}\right)\left|_{z=z_{\IR}}\right.=0\,,
\label{irboundary condition}
\eeq
where the 5D field strenght is defined as $L_{MN}=\partial_M L_N-\partial_N L_M-i[L_M,\,L_N]$ and analogously for $R$.
At the other boundary, the UV-boundary, we impose Dirichlet conditions to all the fields:
\beq
 L_\mu\left|_{z=z_{\UV}}\right.=R_\mu\left|_{z=z_{\UV}}\right.=0\, .
\label{uvboundary condition}
 \eeq
 The Kaluza--Klein (KK) and holographic description of the 5D theory described above
 has been extensively studied in the literature \cite{Csaki:2003zu, Erlich:2005qh}.
From a 4D point of view, the theory resembles to large-$N_c$ QCD
where the Goldstone bosons or pions 
are associated to  the fifth gauge field component, and the $\rho$ and $a_1$
resonances  correspond to the gauge KK-states.
At low-energies all these 5D theories  are described 
by an $(SU(2)_L\times SU(2)_R)/SU(2)_{V}$  non-linear $\sigma$-model.

We are looking for static finite energy solutions of the classical equations of motion (EOM)
of the above 5D theory. 
4D Lorentz invariance allows us to take the anstatz $L_0=R_0=0$, $L_{\hat\mu}=L_{\hat\mu}({\mathbf x},z)$ and $R_{\hat\mu}=R_{\hat\mu}({\mathbf x},z)$, where $0$ and $\hat\mu$ label, respectively, the temporal and spacial coordinates while ${\mathbf x}$ denotes ordinary $3$-space. Starting from the usual 5D YM action the energy of this field configuration reads
\beq
E=\int d^3{x}\int^{z_{\IR}}_{z_{\UV}} dz\,  a(z)\, \frac{M_5}{2} 
\Tr\left[{L_{\hat\mu\hat\nu}L^{\hat\mu\hat\nu}}+{R_{\hat\mu\hat\nu}R^{\hat\mu\hat\nu}}\right]\,,
\label{l5}
\eeq
where the indeces are now raised and lowered by the Euclidean 4D metric. Finding solutions to the EOM is the same as minimizing the energy functional Eq.~(\ref{l5}), which closely resembles the Euclidean YM action in 4D. Our problem is therefore very similar to the one of finding $SU(2)$ instantons, even though, as we will see later, there are some important differences which will not allow us to find an analitic solution.

The topological charge of the soliton will be  defined by 
\beq
Q=\frac1{32\pi^2}\int d^3x\int^{z_{\IR}}_{z_{\UV}} dz\,  
\epsilon_{\hat\mu\hat\nu\hat\rho\hat\sigma}\Tr\left[{L^{\hat\mu\hat\nu}L^{\hat\rho\hat\sigma}}
-{R^{\hat\mu\hat\nu}R^{\hat\rho\hat\sigma}}\right]\,,
\label{topcharge}
\eeq
which is the difference between the $L$ and $R$ instanton charges. 
In order to show that $Q$ is a topological integer number, and with the aim of making the relation with the skyrmion more precise, it is convenient to go to the axial gauge 
$L_5=R_5=0$. The latter can be easily reached, starting from a generic gauge field configuration, by means of a Wilson-line transformation. 
In the axial gauge both boundary conditions Eqs.~(\ref{irboundary condition}) and (\ref{uvboundary condition}) cannot be
simultaneously satisfied. Let us  then 
keep  Eq.~(\ref{irboundary condition})  but modify  the UV-boundary condition to
\beq
{\widetilde L}_{i}\left|_{z=z_{\UV}}\right.=i\,U({\mathbf x})\partial_iU({\mathbf x})^\dagger\,,\;\;\;\;\;{\widetilde R}_{i}\left|_{z=z_{\UV}}\right.=0\,,
\label{uvnboundary condition}
\eeq
where $\widetilde L_i$ and $\widetilde R_i$ are the gauge fields in the axial gauge and 
$i$ runs over the $3$ ordinary space coordinates.  The field $U({\mathbf x})$ in the equation above precisely corresponds to the Goldstone field in the 4D interpretation \cite{Panico:2007qd} once a static Ansatz is taken.
By using a form notation \cite{Chu:1996fr} $A=-i\,A_{\hat\mu} dx^{\hat\mu}$ and 
 remembering that $F\wedge F=d\omega_3(A)$,  the 4D integral in Eq.~(\ref{topcharge}) can be rewritten as an integral of the third Chern--Simons form $\omega_3(A)$ on the $3D$ boundary of the space:
 \beq
Q\,=\,\frac1{8\pi^2}\int_{3D}\left[\omega_3({\widetilde L}) -
\omega_3({\widetilde R})\right]\, .
\label{qw3}
 \eeq
   The contribution to $Q$ coming from  the IR-boundary
 vanishes as the $L$ and $R$ terms in Eq.~(\ref{qw3}) cancel each other due to Eq.~(\ref{irboundary condition}).
  This is crucial for $Q$ to be quantized and it is the reason why we have to choose the relative minus sign among the $L$ and $R$ instanton charges in the definition of $Q$. 
At the ${\mathbf x}^2\rightarrow\infty$ boundary, the contribution to $Q$ also vanishes
since in the axial gauge $\partial_5 A_i=0$ (in order to have $F_{5i}=0$).    
We are then  left with the UV-boundary  which we can topologically regard
as the $3$-sphere $S_3$. Therefore, we find
\beq
Q\,=\,-\frac1{8\pi^2}\int_\UV\omega_3\left[{\widetilde L}_i\left(=i\,U\partial_iU^\dagger\right)
\right]\,=\,\frac1{24\pi^2}\int d^3x\,  \epsilon^{ijk}{\textrm Tr}\left[
U\partial_iU^\dagger\,U\partial_jU^\dagger\,U\partial_kU^\dagger\right]\,\,\in\ZZ\,.
\label{qfinal}
\eeq
The charge $Q$ is equal to the Cartan--Maurer integral invariant for $SU(2)$ which is an integer.
Then  solutions with nonzero $Q$, 
if they exist, cannot trivially correspond to a pure gauge configuration and they must have positive energy. Moreover, the particles associated to solitons with $Q=\pm1$ will be stable given that they have minimal charge.
 Eq.~(\ref{qfinal}) also tells us that 
 the non-trivial configuration    $U({\mathbf x})$ corresponds to a 4D skyrmion
 with $B$ being the  baryon number. 
In  a general gauge, the value of  $U({\mathbf x})$   will be given by
\beq
U({\mathbf x})\,=\,P\left\{\exp\left[-i\,\int_{z_{\UV}}^{z_{\IR}}dz'\,R_5({\mathbf x},z')\right]\right\}\cdot P\left\{\exp\left[i\,\int_{z_{\UV}}^{z_{\IR}}dz'\,L_5({\mathbf x},z')\right]\right\}\, ,
\label{sky}
\eeq
where $P$ indicates   path ordering.  
From  a KK  perspective, the 5D soliton that we are looking for 
can be considered to be a 4D skyrmion made
of Goldstone bosons and  the  massive tower of KK gauge bosons.


We want to  find  a  numerical solution to the 4D EOM arising from  the energy functional of Eq.(\ref{l5}) with suitable boundary conditions enforcing $Q=1$. To do that, as we will now discuss, the axial gauge is not an appropriate choice. We will then restart 
with in  a gauge-independent way and specify later on the new gauge-fixing condition. It is expected on general grounds that the  solution will be maximally symmetric, {\it i.e.} invariant under all symmetry transformations which are compatible with the boundary conditions. We impose, first of all, invariance under ''cylindrical'' transformations, {\it i.e.} combined $SU(2)$ gauge and $3D$ spacial rotations. This corresponds 
to the following Ansatz \cite{Witten:1976ck}: 
\bea
L^a_j&=&-\frac{1+\phi_2^{L}(r,z)}{r^2}\, \epsilon_{jak}{x}_k+\frac{\phi_1^{L}(r,z)}{r^3}\left(r^2\delta_{ja}-{x}_j{x}_a\right)+\frac{A_1^L(r,z)}{r^2}{ x}_j{ x}_a\,,\nn\\
L^a_5&=&\frac{A^L_2(r,z)}{r}{ x}^a\,,
\label{ans1}
\eea
and similarly for $R_{\hat\mu}$. We have reduced our original 4D problem to a much simpler $2$ dimensional one, being $x^{\bar\mu}=\{r,z\}$ (where $r=\sqrt{{\mathbf x}^2}$) the 2D coordinates. The Ansatz has partially fixed the gauge  leaving only a $U(1)_L\times U(1)_R$ subgroup. The allowed transformations are those which preserve the cylindrical symmetry and have the form $g_{L,R}={\textrm{exp}}[i\,\alpha_{L,R}(r,z){x}^a\sigma_a/(2r)]$. Under this symmetry $A_{\bar\mu}^{L,R}$ transform as gauge fields ($A_{\bar\mu}^{L,R}\rightarrow A_{\bar\mu}^{L,R}+\partial_{\bar\mu}\alpha_{L,R}$) and $\phi_{L,R}=\phi^{L,R}_{1}+i{\phi_2}^{L,R}$ are complex scalars of charge $+1$.

In addition to the global symmetries, our problem  has also discrete ones. Those are the $L\leftrightarrow R$ interchange and ordinary parity ${\mathbf x}\rightarrow -{\mathbf x}$. The charge $Q$ is odd under each of these transformations, so
 we cannot impose our solution to be separately invariant under both, but only under their combined action. We can then further reduce our Ansatz by imposing $L_{\hat\mu}(\mathbf x,z)=R_{\hat\mu}(-\mathbf x,z)$:
\bea
&A_1\equiv A_1^R=-A_1^L\,,\;\;\;\;\; &A_2\equiv A_2^R=-A_2^L\,,\nn\\
&\phi_1\equiv\phi^R_1=-\phi^L_1\,,\;\;\;\;\; &\phi_2\equiv\phi^R_2=\phi^L_2\,.
\label{ans2}
\eea
Our solution is now fully specified by $4$ real 2D functions $A_{\bar\mu}$ and $\phi$. We have a residual $U(1)$ invariance corresponding to $g_L^\dagger=g_R={\textrm{exp}}[i\,\alpha(r,z)x^a\sigma_a/(2r)]$ under which $A_{\bar\mu}$ is the gauge field and $\phi$ has charge $+1$. 

By  substituting  the Ansatz Eqs.~(\ref{ans1}) and (\ref{ans2}) into the energy Eq.~(\ref{l5}), one finds
\beq
E=16\pi\int_0^\infty dr\int^{z_{\IR}}_{z_\UV}dz\,M_5\,a(z)\left[
\frac{1}{2}|D_{\bar\mu}\phi|^2+\frac{1}{8}r^2F^{2}_{\bar\mu\bar\nu}
+\frac{1}{4r^2}\left(1-|\phi|^2\right)^2\right]\,.
\label{en2d}
\eeq
In   flat space, $a(z)=1$, this corresponds to a  2D Abelian Higgs model 
(with metric $g^{\bar\mu\bar\nu}=r^2\delta^{\bar\mu\bar\nu}$). 
For a general warp factor we have, however,  a non-metrical theory. 
Substituting the Ansatz in the topological charge Eq.~(\ref{topcharge}) we find
\beq
Q=\frac{1}{2\pi}\int_{0}^\infty dr\int^{z_{\IR}}_{z_{\UV}}dz\,\epsilon^{\bar\mu\bar\nu}
\bigg[\partial_{\bar\mu}(-i \phi^*D_{\bar\nu}\phi+h.c.)
+F_{\bar\mu\bar\nu}\bigg]\,.
\label{topcharge2d}
\eeq
The charge can be written, as it should, as an integral over the $1D$ boundary of the 2D space. We will choose the boundary conditions in such a way that the first term 
of Eq.~(\ref{topcharge2d}) vanishes, and therefore $Q$  will  coincide with the magnetic flux, {\it i.e.} the topological charge of the Abelian Higgs model.

In order to solve  numerically  the EOM associated with Eq.~(\ref{en2d}),
 they must be recast in the form of a 2D system of non-linear Elliptic Partial Differential Equations (EPDE). The numerical resolution of the 2D EPDE boundary value problem has indeed been widely studied and very simple and powerful packages exist. This is the reason why we cannot  work in the axial gauge, since   the EOM  for $A_1$ is not   elliptic
 in this gauge. 
 We will impose the 2D Lorentz gauge condition $\partial_{\bar\mu}A^{\bar\mu}=0$. The equations for $A_{\bar\nu}$ become  $J^{\bar\nu}=\partial_{\bar\mu}(r^2a\,F^{\bar\mu\bar\nu})=r^2a\,\Box A^{\bar\nu}+\partial_{\bar\mu}(r^2a)F^{\bar\mu\bar\nu}$, where $\Box$ is the 
 2D Laplacian and $J$ is the current of the field $\phi$. In this way we have a system of $4$ EPDE and $4$ real unknown functions which we can determine numerically. 
 Since we
 are  counting the two gauge field components  
  as independent functions,  one can wonder
  whether   the gauge-fixing condition is satisfied.
 Notice however that   
 by taking the $\partial_{\bar\nu}$ derivative of the EOM for $A_{\bar\nu}$ and observing that the current is conserved, $\partial_{\bar\nu}J^{\bar\nu}=0$, on the solutions of the EOM for $\phi$, we find an EPDE for $\partial_{\bar\nu}A^{\bar\nu}$ which reads $[(r^2a)\Box+\partial_{\bar\mu}(r^2a)\partial^{\bar\mu}](\partial_{\bar\nu}A^{\bar\nu})=0$. This equation has a unique solution once the boundary conditions are given. If we impose $\partial_{\bar\nu}A^{\bar\nu}=0$ at the boundaries, the gauge will then be automatically mantained everywhere in the bulk.

We will solve the EOM in the rectangle $\left(z\in[z_\UV,z_\IR]\,,\;r\in[0,R]\right)$ and we will numerically take the limit $R\rightarrow\infty$. At the three sides $z=z_\IR$, $z=z_\UV$ and $r=R$ we impose the following boundary conditions
\beq
z=z_\IR\,:\,\left\{\begin{array}{l}\phi_1=0\\\partial_2\phi_2=0\\A_1=0\\\partial_2A_2=0\end{array}\right.\, ,\;\;\;\;\;
z=z_\UV\,:\,\left\{\begin{array}{l}\phi_1=0\\\phi_2=-1\\A_1=0\\\partial_2A_2=0\end{array}\right.\, ,\;\;\;\;\;\;
r=R\,:\,\left\{\begin{array}{l}\phi=-i\, e^{i\pi z/L}
\\\partial_1A_1=0\\A_2=\frac{\pi}{L}
\end{array}\right.\, ,
\label{boundary condition}
\eeq
where $L$  is the conformal length of the extra dimension:
 \beq
 L=\int_{z_\UV}^{z_\IR} dz=z_\IR-z_\UV\, .
 \eeq
 Few comments  are in order.
The conditions on the IR and UV boundary for $\phi_{1,2}$ and $A_1$ come 
respectively from
 Eqs.~(\ref{irboundary condition}) and (\ref{uvboundary condition}), while the condition for $A_2$ arises from
 the gauge fixing.  
  On the boundary at $r=R$ we have chosen  the fields $\phi$ and $A_2$
with   a nontrivial
  profile  consistent, however, with $E\rightarrow 0$ at $R\rightarrow\infty$;
  the condition for $A_1$ comes again from the gauge fixing.
  Our choice of   boundary conditions  is such that  the charge Eq.~(\ref{topcharge2d})
receives contributions only from the boundary at $r=R$.
The $r=0$ boundary  of our rectangular domain is special, given that the EOM become  singular there. The software which we will employ (FEMLAB 3.1) permits us to extend the domain up to $r=0$ because it never uses the EOM on the boundary lines; the equations on that points are provided by the boundary conditions. Some special care is required, however, to enforce the program to converge to a regular solution. We find that to impose regularity the following conditions are needed: 
\beq
r=0\,:\,\left\{\begin{array}{l}\phi_1/r\rightarrow A_1\\\
(1+\phi_2)/r\rightarrow 0\\\
A_2\rightarrow 0\\\
\partial_1A_1=0
\end{array}\right.\, .\label{boundary conditionr0}
\eeq
 The conditions for $\phi_{1,2}$  and $A_2$ are   extracted from Eq.~(\ref{ans1}) 
 by requiring the gauge fields to be well defined 5D vectors, 
 while the  condition for $A_1$ is the gauge fixing.
 To impose these conditions and obtain regular solutions we define the
 rescaled fields $\chi_{1,2}$:  $\phi_1=r\chi_1$ and  $\phi_2=-1+r\chi_2$.
  The boundary condition Eq.~(\ref{boundary conditionr0})  reads now $\chi_1=A_1$, $\chi_2=0$ and $A_2=0$, plus  the gauge-fixing condition $\partial_1A_1=0$. 
The rescaled fields $\chi_{1,2}$, together with $A_{1,2}$,
are actually the ones used in 
 our numerical computations.
 
Any gauge-invariant information about the 5D soliton such 
as the energy  and charge densities Eqs.~(\ref{en2d}) and (\ref{topcharge2d}) respectively
can be directly extracted with our procedure. 
It would be also interesting, however, to know also what the profile of the skyrmion is. To this end we have to go to the axial gauge, as we discussed above. From  Eq.~(\ref{sky}) we have
\be
U(\mathbf x)=\exp\left[-i\,\sigma_ix^i/r\int_{z_\UV}^{z_\IR}dz'A_2(r,z')\right]
\,\equiv\,\exp\left[i\,f(r)\sigma_i x^i/r\right]
\, .
\ee
We immediatly see that our boundary conditions imply $f(0)=0$ and  $f(\infty)=-\pi$, 
as it should be  for  a skyrmion in which  $f(r)$ undergoes a $-\pi$ variation when $r$ goes from $0$ to $\infty$.

When applying the above described numerical method, however, we do not find any solution for either a flat or warped  extra dimension. The 5D theory which we are considering does not possess any non-singular soliton and 
%
the reason is the following \cite{Hong:2007kx,
Hata:2007mb}.
The energy Eq.~(\ref{l5}) of a 5D soliton configuration 
 is bounded from below: 
\beq
E\,\geq 
\int d^3x\int^{z_{\IR}}_{z_{\UV}} dz\,a(z)\frac{M_5}{4}\Tr
\label{bound}
\left|
\epsilon_{\hat\mu\hat\nu\hat\rho\hat\sigma}{F^{\hat\mu\hat\nu}F^{\hat\rho\hat\sigma}}
\right|\,
\geq\, 8\pi^2M_5\left|Q\right|\, ,
\eeq
where we take taken, for simplicity, the Ansatz  
$L_{\hat\mu\hat\nu}(\mathbf x,z)=R_{\hat\mu\hat\nu}(-\mathbf x,z)\equiv F_{\hat\mu\hat\nu}(\mathbf x,z)$
and 
in the last inequality we have used  $a(z)\geq a(z_{\IR})=1$.
The first inequality can be  saturated if  $F_{\hat\mu\hat\nu}$  is self-dual or antiself-dual,
while the 
 the lower bound 
$8\pi^2M_5|Q|$ 
is  approached   (in warped spaces) 
  when these (anti)self-dual configurations
are centered at  $z=z_\IR$ and tend to zero size (since $a(z)$ has its minimum
at  $z=z_\IR$).
This is the case of  a 4D YM instanton configuration of 
 infinitesimal  size, $\rho\rightarrow0$, 
   with the identification of the Euclidean time  $t_E$ with $z$, 
 and  centered at  $z=z_{\rm IR}$.  
 This is clear since the limit $\rho\rightarrow0$ is equivalent to
 keeping  the instanton size fixed  
 and taking to zero all other dimensional parameters, the curvature and the extra dimensional
 size
,  $a(z)\rightarrow 1$ and  $z_\UV\rightarrow -\infty$ respectively.
 In this limit the energy   Eq.~(\ref{l5}) corresponds to  the 4D Euclidian action  
that is minimized by an instanton of arbitrary size.
 This means that  the 5D soliton  that we are looking for  is a singular instanton configuration.

 An alternative way to see   
 this  is by     explicitly calculating   the energy of the instanton configuration
in the limit of small size.
This is done in the  Appendix.
For a flat space we obtain
\beq
E(\rho)\,\simeq\,8\pi^2M_5\left[1+\frac14\left(\frac\rho{L}\right)^4\right]\, ,
\label{flaten}
\eeq
while for the AdS space we have 
\beq
E(\rho)\,\simeq\,8\pi^2M_5\left[1+\frac{\rho}{2L}\right]\, .
\label{adsen}
\eeq
In both cases we see that the minimum of the energy is achieved for $\rho\rightarrow 0$.

\subsection{IR-boundary terms}

The problem of stabilizing the radius of the soliton configuration
can be overcome if there is a   repulsive  force  on  the IR-boundary 
that pushes the instanton to large sizes. 
The simplest possibility is to introduce  an IR-boundary kinetic term for the gauge fields:
\beq
S_{\rm IR}=-\int d^4x\,  \, \frac{\alpha}{2} 
\Tr\left[{L_{\mu\nu}L^{\mu\nu}}+{R_{\mu\nu}R^{\mu\nu}}\right]\Big|_{z=z_{\IR}}\, .
\label{l4}
\eeq
The presence of this term is expected in these  theories since 
it is unavoidably induced at the loop level. 
An NDA estimate gives $\alpha\sim 1/(16\pi^2)$. 
Furthermore,  as we will show in section~\ref{calc},
Eq.~(\ref{l4}) 
 is  the lowest higher-dimensional operator of our effective 5D theory and therefore the most
important one beyond Eq.~(\ref{l5}).

Using Eqs.~(\ref{ans1}) and (\ref{ans2}) and  the fact that $\phi_1=A_1=0$ at $z=z_{\IR}$, 
we can write the boundary energy Eq.~(\ref{l4}) as
\beq
E_{\rm IR}=8\pi\alpha\int^\infty_0 dr \left[(\partial_1\phi_2)^2+\frac{1}{2r^2}(1-\phi_2^2)^2
\right]\Bigg|_{z=z_{\IR}}\, .
\label{eboun}
\eeq
We can see that this term  favors large instanton configurations 
by substituting the instanton Eq.~(\ref{inst}) in  Eq.~(\ref{eboun}). We obtain
\beq
E_{\rm IR}(\rho)=6\pi^2\frac{\alpha}{\rho}\,,
\label{ebndy}
\eeq
that grows for small $\rho$.
Therefore   we expect that the presence of Eq.~(\ref{eboun}) will guarantee  the existence of
  non-singular 5D solitons.
These configurations cannot be found analytically and for this reason we have to rely on
numerical  analysis.
The  effect of Eq.~(\ref{eboun})  corresponds  to  a  
 change of the  IR-boundary condition  for $\phi_2$.
 Instead of    that in  Eq.~(\ref{boundary condition}), we have now
\begin{figure}[t]
\ 
\vspace{-1.5cm}
\begin{center}
\epsfig{file=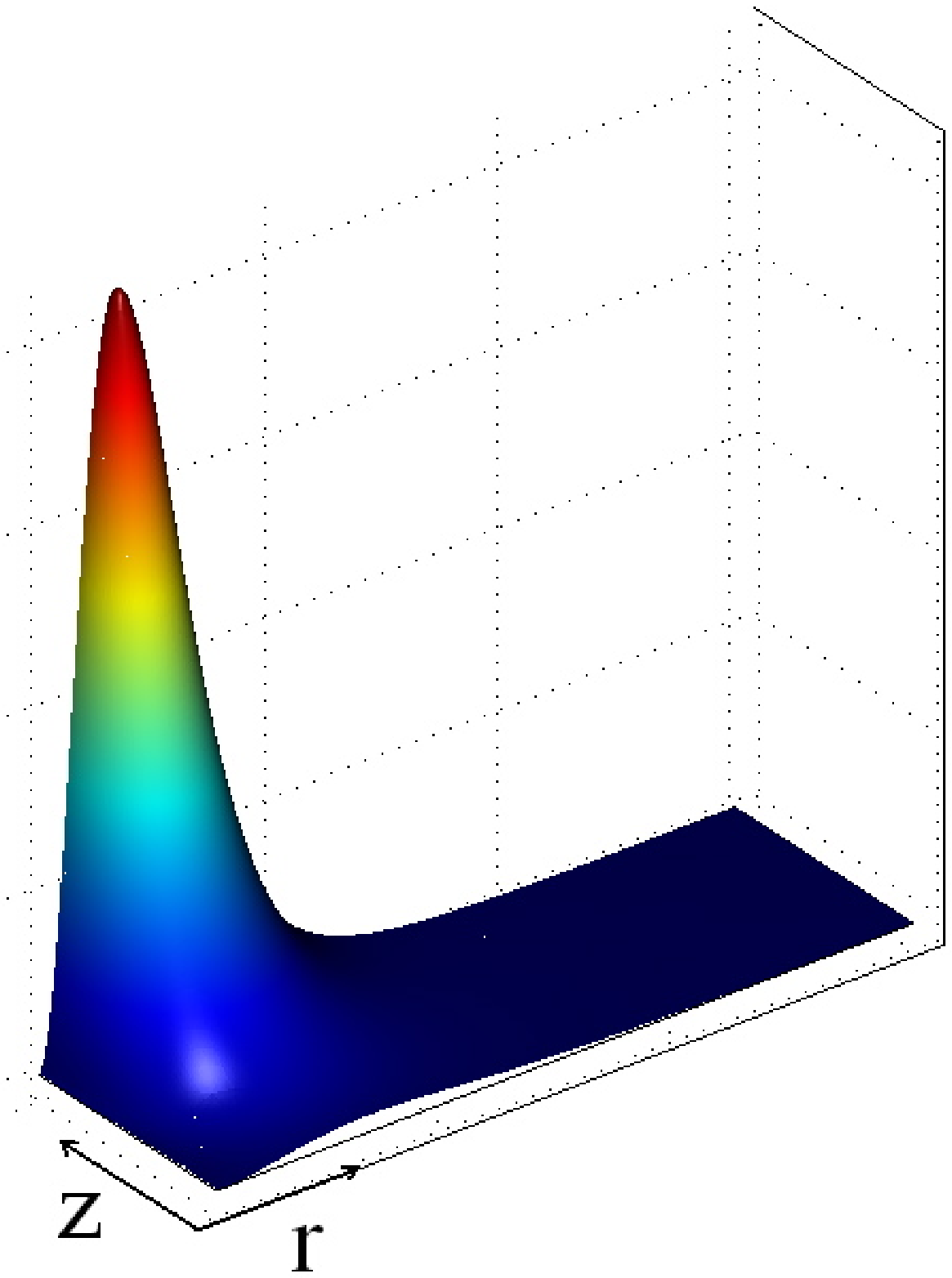,width=0.4\textwidth,angle=0}
\hspace{0.5cm}
\epsfig{file=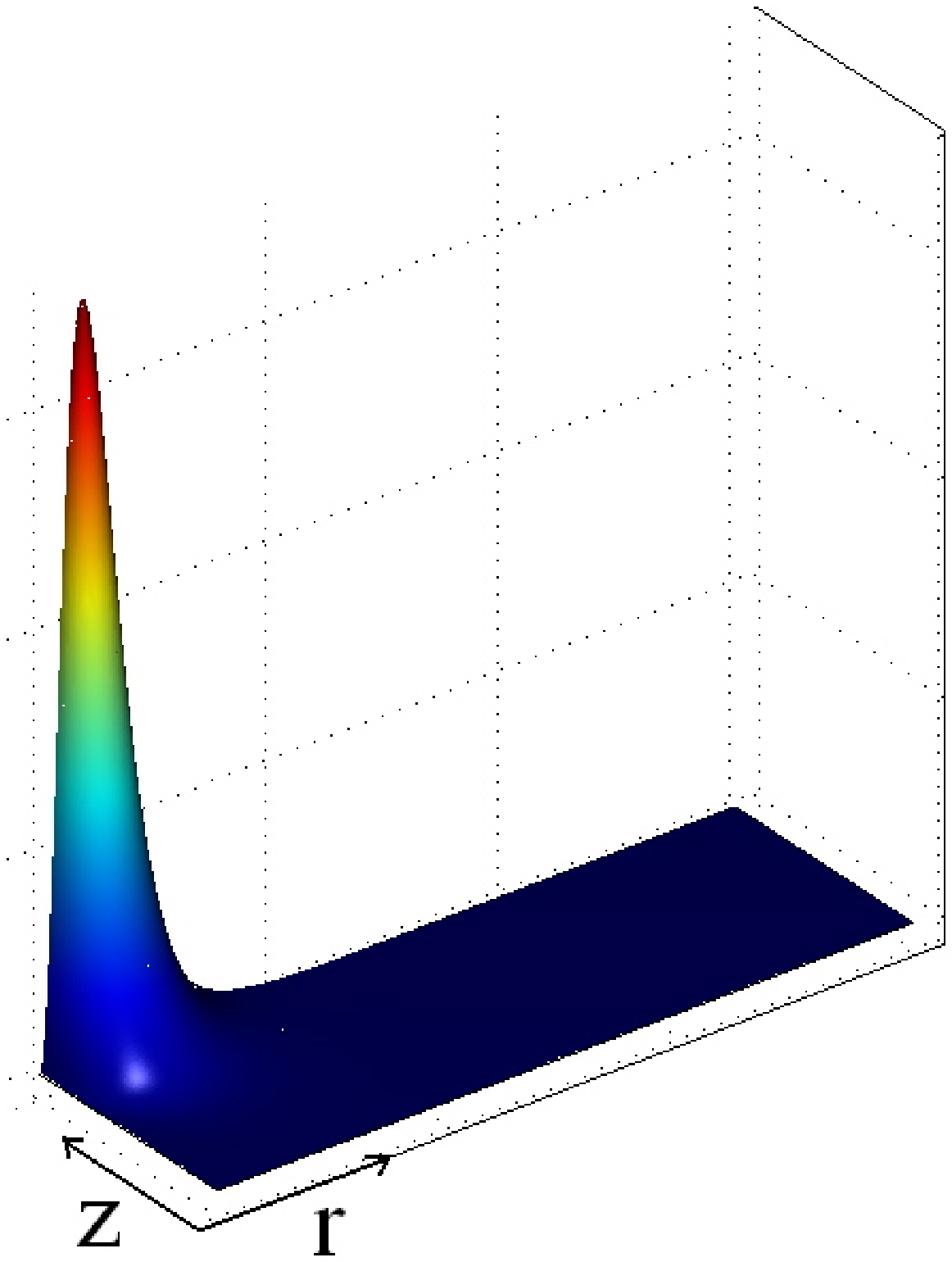,width=0.4\textwidth,angle=0}
\ 
\end{center}
\caption{We plot the bulk energy density Eq.~(\ref{en2d}) of our numerical solution for flat (left panel) and AdS (right panel) space. Both plots are obtained for $\alpha/(L M_5)=0.1$.
}
\label{bolaplot}
\end{figure}
\begin{figure}[t]
\begin{center}
\epsfig{file=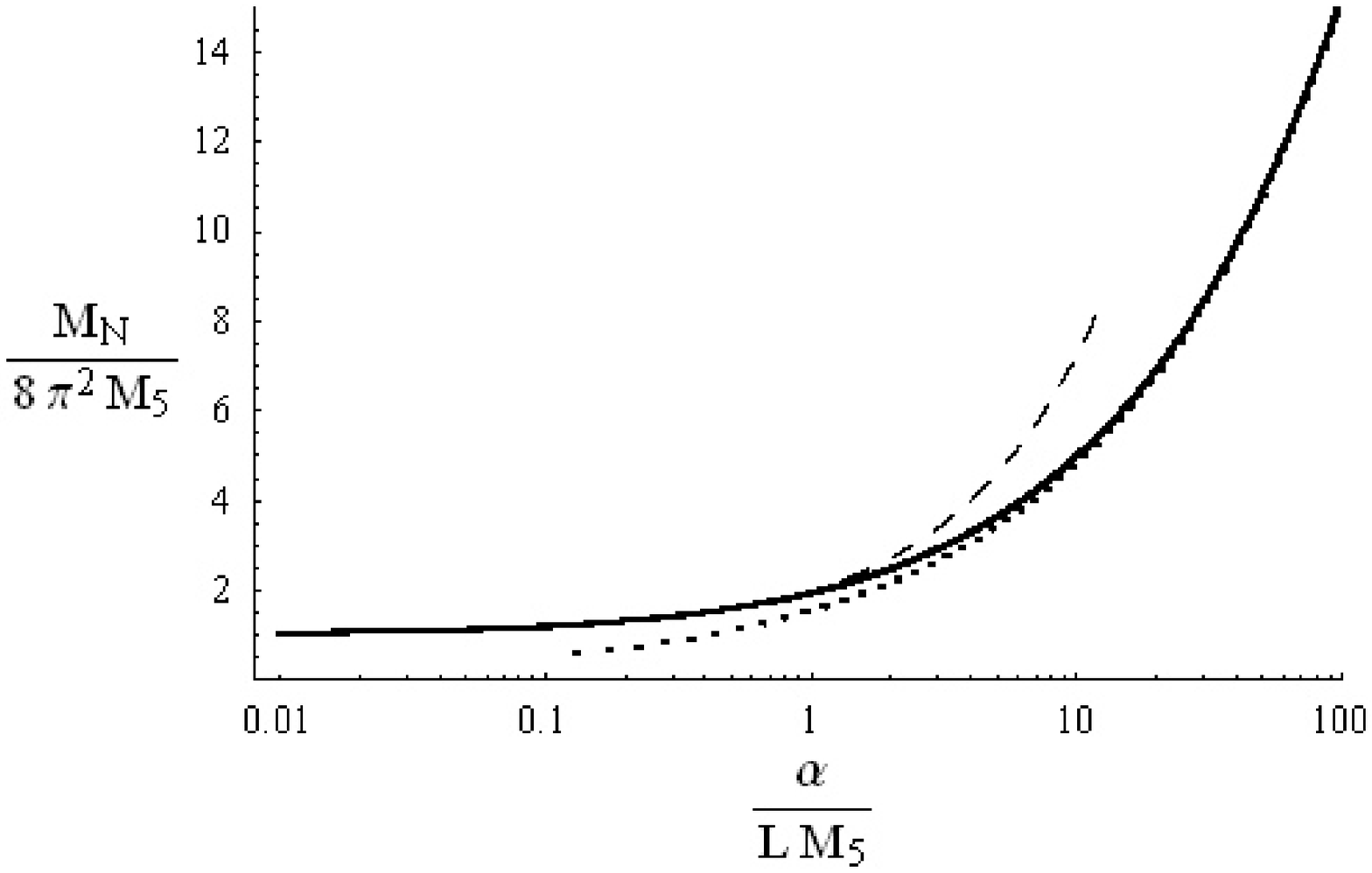,width=0.45\textwidth,angle=0}
\epsfig{file=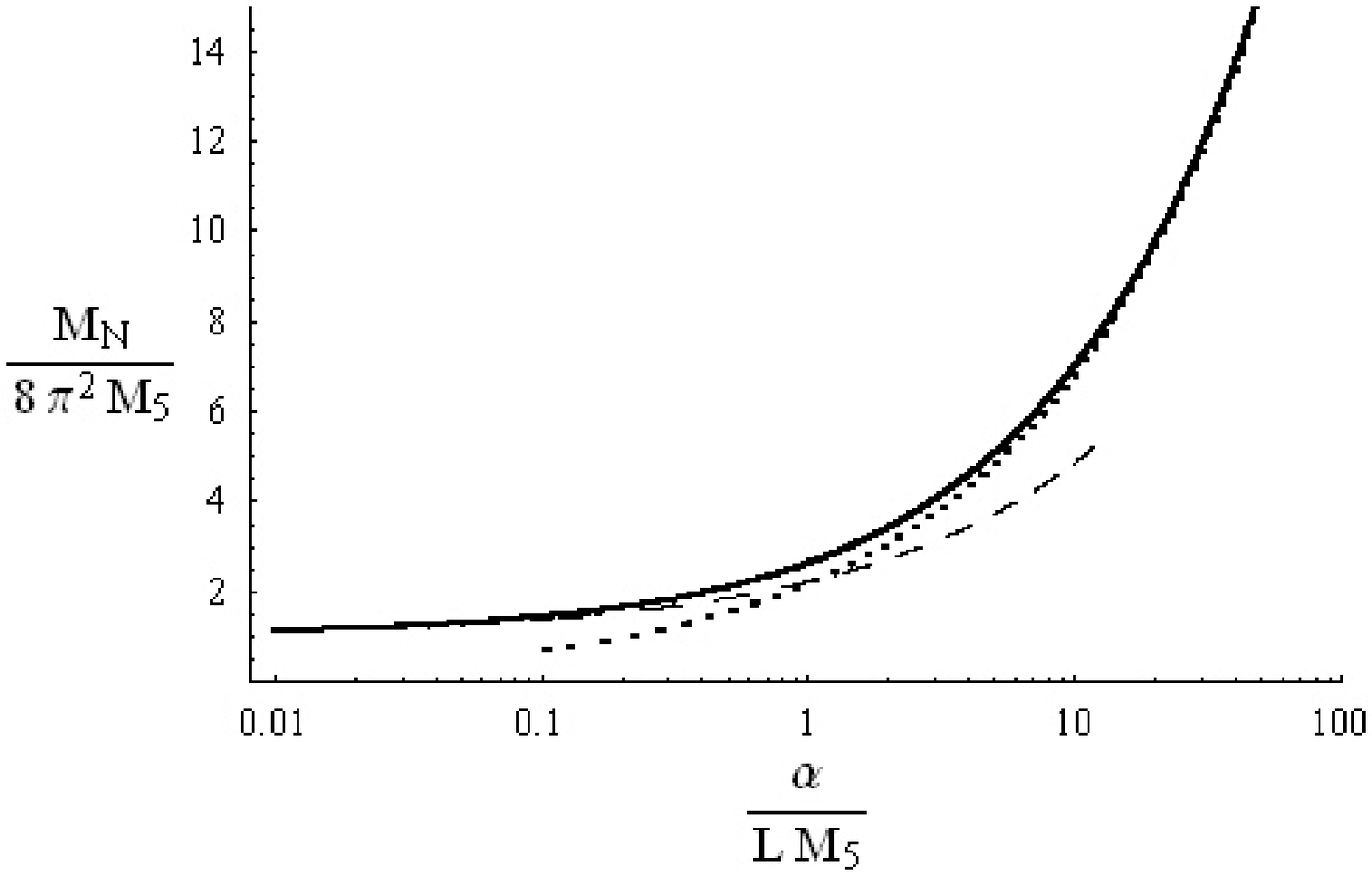,width=0.45\textwidth,angle=0}
\vspace{-.7cm}
\end{center}
\caption{Left panel: the thick line represents the soliton energy which we obtained numerically in the case of flat space. This is well approximated 
by Eq.~(\ref{etfl}) (dashed line) and Eq.~(\ref{massba}) (dotted line) for, respectively, small and large values of $\alpha/(L M_5)$. Right panel: the same plot for AdS; the approximated curves are provided by Eq.~(\ref{radiusads}) (dashed line) and Eq.~(\ref{massba}) (dotted line).
}
\label{energyplot}
\end{figure}
\beq
\partial_2 \phi_2\Big|_{z=z_{\IR}}=\frac{\alpha}{M_5}\left.\left[\partial_1^2\phi_2+\frac{1}{r^2}\phi_2(1-\phi^2_2)\right]\right|_{z=z_{\IR}}\, .
\eeq

\subsection{Numerical results}

To find   numerically the 5D soliton  configuration we use the 
  software package FEMLAB 3.1 \cite{comsol}. 
  This   is  a numerical package for solving  EPDE based on the 
finite element method. 
We have  concentrated on two different spaces: 
\bea
\text{Flat:}&&  a(z)=1\, ,\\  
\text{AdS:}&&  a(z)=\frac{z_\IR}{z}\ \ \text{with}\ \ z_\UV\rightarrow0\, .\label{adsmetric} 
\eea
Fig.~\ref{bolaplot} shows an example of the numerical results that  we get;
for $\alpha/(L M_5)=0.1$ we plot the bulk energy density of the soliton in flat and AdS space. Notice that the size of the AdS soliton is sensibly smaller than the flat one and that the energy density vanishes at $z=z_\UV$  in the case of AdS while it does not in flat space. This is consistent with the holographic CFT interpretation of the AdS soliton.  
Being localized at the IR-boundary, 
the soliton is   a composite state of the CFT.

\begin{figure}[t]
\begin{center}
\epsfig{file=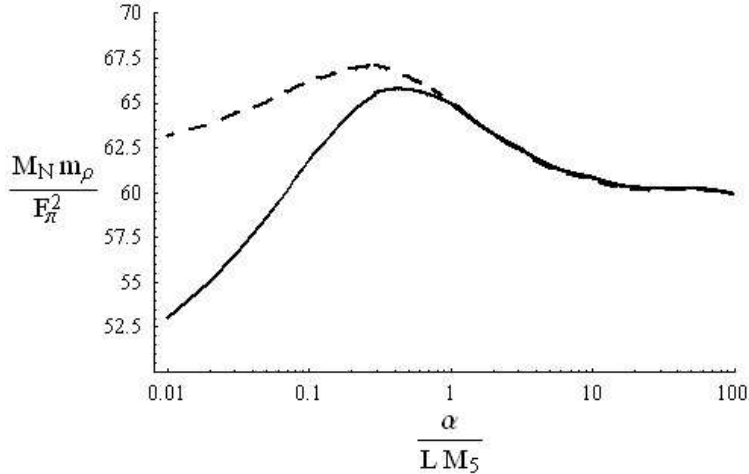,width=0.6\textwidth,angle=0}
\vspace{-.7cm}
\end{center}
\caption{The ratio $M_Nm_\rho/F_\pi^2$ is plotted as a function of $\alpha/(L M_5)$ for a flat(dashed line) and AdS space. The behaviour for large $\alpha/(L M_5)$ is consistent with 
Eq.~(\ref{massba}) while at small $\alpha/(L M_5)$
we obtain Eq.~(\ref{mnas}).}
\label{ratioplot}
\end{figure}

\begin{figure}[t]
\begin{center}
\epsfig{file=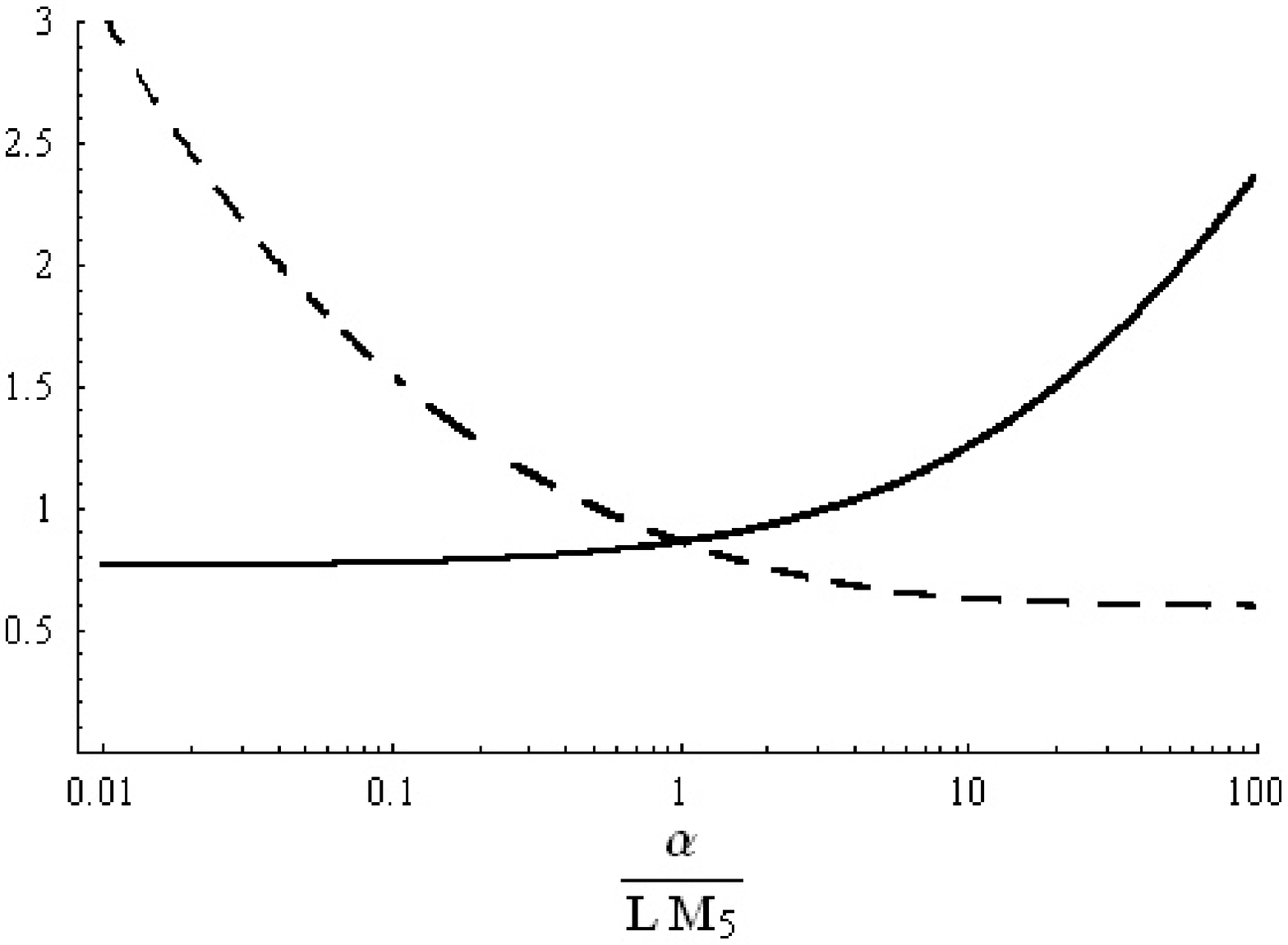,width=0.45\textwidth,angle=0}
\hspace{0.4cm}
\epsfig{file=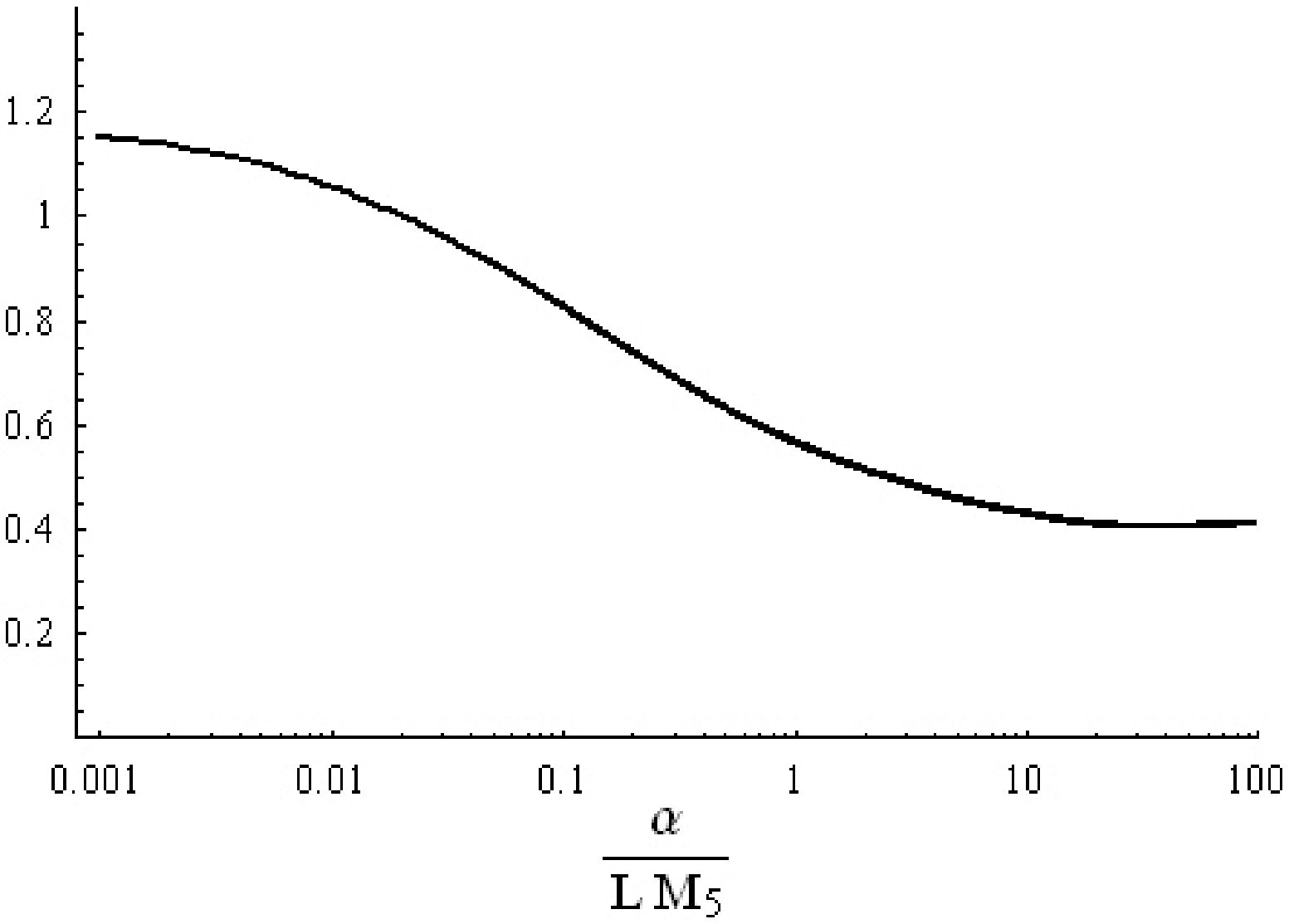,width=0.45\textwidth,angle=0}
\vspace{-.7cm}
\end{center}
\caption{Left panel: $\rho_0[M_5/(\alpha L)]^{1/2}$ (dashed) and $\rho_0[M_5/(\alpha L^4)]^{1/5}$ for the case of flat space. Right panel: $\rho_0[M_5/(\alpha L)]^{1/2}$ in AdS. The limiting behaviour of those quantities is consistent with Eqs.~(\ref{radiusflat}), (\ref{radiusads}) and (\ref{radiuslarge}).}
\label{radioplot}
\end{figure}

In Fig.~\ref{energyplot} we present  the total energy  of the soliton configuration  $E_{total}=E+E_{\rm IR}$
for different  values of  $\alpha/(M_5L)$.
This approximately corresponds to  the mass of the soliton
\footnote{Here we work at the semi-classical level neglecting  corrections due to the quantization of the soliton.}
 $M_N\simeq E_{total}$. 
Although the energy of the soliton  is quite sensitive to $\alpha/(M_5L)$ 
for large values of this quantity,
this is not the case when $E_{total}$ 
is expressed 
as a function of the PGB decay constant $F_\pi$
and the mass of the lowest KK-state $m_\rho$ that are calculated to be
\bea
&& F_\pi^2=\frac{2M_5}{L} ,\quad\quad 
m_\rho\simeq \frac{\pi}{2L}\left(1+\frac{\pi^2\alpha}{4M_5L}\right)^{-1/2}\quad\quad \text{for flat space}\, ,\label{kkmassf}\\
&& F_\pi^2=\frac{4M_5}{L} ,\quad\quad m_\rho\simeq \frac{3\pi}{4L}\left(1+
\frac{9\pi^2\alpha}{32M_5L}\right)^{-1/2}\quad\quad\text{for AdS space}\, ,\label{kkmassa}
\eea
where $F_\pi$ is 
  normalized such that the coefficient 
  in front of the kinetic term of the Goldstone  is $F^2_\pi/4$.
In Fig.~\ref{ratioplot} we show the ratio $M_N{m_\rho}/{F^2_\pi}$ for different values of 
$\alpha$. We see that this ratio  only varies a  $20\%$   
and stays in the range $50-65$,
resulting  in a relatively  model-independent prediction of the soliton mass.
We define the radius $\rho_0$ of the configuration as the mean radius of the
 topological charge density, {\it i.e.}
\beq
\rho_0^2\,=\,\frac{1}{2\pi}\int_{0}^\infty
dr\int^{z_{\IR}}_{z_{\UV}}dz\,\frac{r^2+(z-z_\IR)^2}2\,\epsilon^{\bar\mu\bar\nu}
\bigg[\partial_{\bar\mu}(-i \phi^*D_{\bar\nu}\phi+h.c.)
+F_{\bar\mu\bar\nu}\bigg]\,.
\eeq
 For the instanton configuration in 
 a flat uncompactified space
$\rho_0$ coincides with $\rho$. The radius $\rho_0$ will go to zero as $\alpha\rightarrow0$ and will diverge for $\alpha\rightarrow\infty$.
In order to  show the behaviour of $\rho_0$ in these two limits,
  we plot in Fig.~\ref{radioplot} 
  the following combinations: $\rho_0[M_5/(\alpha L)]^{1/2}$ for both flat and AdS space and $\rho_0[M_5/(\alpha L^4)]^{1/5}$ for flat space only. 
  We will show below how these behaviours can be analytically deduced.

The numerical results presented above have a simple interpretation
 in the limit in which  the boundary term    Eq.~(\ref{l4})  dominates or not  over 
  the bulk   Eq.~(\ref{l5}). 
This corresponds to the limits
 $\alpha\gg M_5L$ and $\alpha\ll M_5L$ respectively. We will discuss  each in turn.

\noindent\underline{ $\alpha\ll M_5L$:}
In this limit  the energy of the soliton  is dominated by the bulk contribution,
and  therefore we expect that the instanton configuration is   a good  approximation of the true solution. 
In this case we can find the size of the configuration by
minimizing the instanton energy with respect to $\rho$.
For flat space,  the minimum of   Eqs.~(\ref{flaten}) and (\ref{ebndy})
corresponds to 
\beq
\rho_0\simeq  \left(\frac{3\alpha}{4M_5L}\right)^{1/5}L\, ,
\label{radiusflat}
\eeq
that leads to  the total energy
\beq
E_{total}\simeq 8\pi^2M_5\left(1+\frac{5}{4}
\left(\frac{3\alpha}{4M_5L}\right)^{4/5}\right)\, .
\label{etfl}
\eeq
For AdS space, the minimization of 
Eqs.~(\ref{adsen})  and (\ref{ebndy}) gives
\beq
\rho_0\simeq \sqrt{\frac{3\alpha}{2M_5L}}\,L\ ,\quad\qquad 
E_{total}\simeq8\pi^2M_5\left(1+\sqrt{\frac{3\alpha}{2M_5L}}\right)\, .
\label{radiusads}
\eeq
These analytical results agree  in the small $\alpha$ limit with the numerical ones.  
From   Fig.~\ref{radioplot} we see in fact that $\rho_0$  scales as $(\alpha/(M_5L))^{1/5}$ 
and $(\alpha/(M_5L))^{1/2}$  for flat and AdS space
respectively  in the limit of small $\alpha$.
The energy in this limit is approximately given by the instanton energy $8\pi^2M_5$ 
independently of the metric or the compactification of the 
extra dimension as can be seen from Fig.~\ref{energyplot}.
Finally, the energy can be rewritten as
\beq
M_N\simeq  2\pi^3\frac{F^2_\pi}{m_\rho}\ \ \ \    (\text{flat space}) ,\quad \ 
M_N \simeq \frac{3\pi^3}{2} \frac{F^2_\pi}{m_\rho}\ \ \ \    (\text{AdS space})\, ,
\label{mnas}
\eeq
in accordance with Fig.~\ref{ratioplot}.

\noindent\underline{ $\alpha\gg M_5L$:}
In this limit the boundary  term Eq.~(\ref{l4}) is dominat,
and one finds that 
the  effective low-energy theory  below $1/L$ 
corresponds to the following one.
It is a 4D theory with a non-linearly realized 
$SU(2)_L\times SU(2)_R$ global symmetry in which 
the $SU(2)_{V}$ subgroup is weakly gauged.
The gauge coupling is given by  $g=1/\sqrt{\alpha}$ 
and the mass of the  gauge boson  is then  $m_\rho=gF_\pi$
where $F_\pi$ is the Goldstone decay constant  defined above.  
This is   valid independently of the geometry of the extra dimension. 
One can find this result by ordinary KK reduction
that below $1/L$ leads to  massless  scalars $(L_5-R_5)$, the Goldstones,
plus a   KK-state of   $(L_\mu+R_\mu)$ with  mass $\sim\sqrt{M_5/(L\alpha)}$
--see Eqs.~(\ref{kkmassf}) and (\ref{kkmassa}).
A better  understanding  of the large $\alpha$ limit, however, 
can be obtained  by following
 a  "holographic"  prescription and    treating  
    the bulk gauge fields  and  its  value at the IR-boundary  as  
 distinct  variables.
The bulk plus the UV-boundary  sector has a global  
$SU(2)_L\times SU(2)_R$ symmetry broken down to nothing due to the boundary condition Eq.~(\ref{uvboundary condition}).  Its 4D spectrum  contains the heavy  states of mass of order $1/L$ and the Goldstone bosons.   
On the other hand,  the   IR-boundary (see Eq.~(\ref{irboundary condition}) and Eq.~(\ref{l4}))  
corresponds to a gauging of only the  
$SU(2)_{V}$  subgroup of the global $SU(2)_L\times SU(2)_R$ with   $1/g^2=\alpha$. 
Therefore the spectrum of the full 5D theory below $1/L$ consists 
of the Goldstone bosons 
parametrizing the  coset $(SU(2)_L\times SU(2)_R)/SU(2)_V$ 
with decay constant $F_\pi$ plus the 
$SU(2)_V$ gauge bosons that  acquire a  mass $gF_\pi\ll 1/L$.
Theories like these were considered in the past 
and  it was shown that they 
 have stable non-singular skyrmions \cite{Igarashi:1985et}. 
The mass of the $Q=1$ skyrmion was found to be 
\footnote{We are taking the result from Ref.~\cite{Igarashi:1985et}
where for   $a=m^2_\rho/(g^2F^2_\pi)=1$,
$m_\rho\simeq 770$ GeV and $F_\pi\simeq  93$ GeV they obtain
 $ M_N\simeq 667$ GeV.} 
\beq
M_N\simeq  59.4\, \frac{F_\pi^2}{m_\rho}\, .
\label{massba}
\eeq
In Fig.~\ref{ratioplot} one can see that our numerical value  of $M_N$ tends to Eq.~(\ref{massba}) 
for large values of $\alpha$.
We also see from  Fig.~\ref{radioplot}   
that the radius  of our 5D soliton scales in the large $\alpha$ limit  as 
\beq
\rho_0\sim\sqrt{\alpha L/M_5}\sim 1/m_\rho\,,
\label{radiuslarge}
\eeq
as expected for a 4D skyrmion.
Thus, we can conclude  that  for $\alpha\gg M_5L$  the  5D soliton  corresponds  to a 4D  skyrmion made of   Goldstones and 
a massive  gauge field.

Summarizing, by ranging $\alpha$ from $0$ to $\infty$,
our 5D  soliton configuration goes
 from being a small  4D instanton configuration  to the  4D skyrmion 
 of Ref.~\cite{Igarashi:1985et}.
By increasing $\alpha$ we are  effectively  decreasing  the contribution to the soliton
of the  heaviest gauge 
KK-states  that reduces to a single KK in the 
   limit of very large $\alpha$. 
   
\subsection{Calculability in the 5D effective theory}
 \label{calc}
 
We have found that once we include the IR term Eq.~(\ref{l4}) into the action a regular static solution to the 5D EOM with $Q=1$ exists. In the spirit of effective field theories, however, the action also contains an infinite number of 5D higher-dimensional operators which are suppressed by inverse powers of the cut-off scale $\Lambda_5$, 
and it is of crucial importance to establish how much our results are sensitive to such operators. 

The contribution to the soliton energy of the higher-dimensional operators is easily estimated by substituting the soliton configuration into the operators. 
Since  these  operators have at least one more dimension of energy,  their  coefficients are suppressed by one more power of $\Lambda_5^{-1}$. 
On dimensional grounds, therefore, the contribution to the energy of the new operators carries on more powers of $1/(\Lambda_5\rho_0)$ where the radius $\rho_0$ is the typical size of the soliton. 
We must  check that  $1/(\Lambda_5\rho_0)\ll 1$  in such a way that a sensible perturbative expansion, analogous to the $E/\Lambda_5$ expantion for a scattering amplitude at energies $E<\Lambda_5$, can be performed. If this is the case the 5D soliton can be studied 
with our 5D effective theory in the sense that we can compute its properties, to a certain degree of accuracy, by only including a finite number of operators into the action. 

By looking  at Eqs.~(\ref{radiusflat}) and (\ref{radiusads}) ,  we find 
that   $\rho_0\Lambda_5\sim(\Lambda_5 L)^{4/5}$ and  $\rho_0\Lambda_5\sim(\Lambda_5 L)^{1/2}$  respectively for a flat and AdS space.
Therefore,  as long as $\Lambda_5 L\ll 1$ 
(as it should for our extra-dimensional theory to make sense),
 we have $1/(\Lambda_5\rho_0)\ll 1$ as needed 
 in order to trust our
  soliton within the 5D effective approach.
We can see this in more detail by considering a specific $6$-dimensional bulk operator such as $F_{MN}D_RD^RF^{MN}$ that, by the NDA counting, is suppressed by $M_5/\Lambda^2_5$ where
$\Lambda_5\sim 24\pi^3 M_5$.
The contribution of this operator to the soliton energy is of order $\delta E\sim 8\pi^2M_5/(\Lambda_5 \rho_0)^{2}$ that  should be compared with the boundary energy  Eq.~(\ref{ebndy}). The relative correction 
$\delta E/E_\IR(\rho_0)\sim(24\pi^3\alpha\Lambda_5\rho_0)^{-1}$ 
 plays the role of the expansion parameter. Using Eqs.~(\ref{radiusflat}) and (\ref{radiusads}) we get, respectively
\beq
\left(24\pi^3\alpha \right)^{-6/5}\left(\Lambda_5 L \right)^{-4/5}\;\;\;\textrm{and}\;\;\;
\left(24\pi^3\alpha \right)^{-3/2}\left(\Lambda_5 L \right)^{-1/2}\,.
\label{conds}
\eeq
If $\alpha$, whose NDA value is $1/(16\pi^2)$, is not unnaturaly small and  
$\Lambda_5 L>1$,  the contribution to the energy of the new operator is safely small. Notice that, contrary to the naive expectation, we did not loose too much 
in our perturbation  parameter because of the fact that we needed to include the operator 
Eq.~(\ref{l4}) to stabilize the soliton. If we would have found a solution for $\alpha=0$, its size would have been given by $\rho_0\sim L$ and we wolud have obtained, instead of Eq.~(\ref{conds}), a perturbative expansion parameter of order $1/(\Lambda_5 L)$. 

We can safely neglect any $d$-dimensional bulk and brane-localized operators with, respectively, $d\geqslant 6$ and $d\geqslant 5$. The only operator that could  be important in our analysis is the five-dimensional Chern-Simons (CS) term which, by NDA,  is suppressed 
by $M_5/\Lambda_5\sim \alpha$, and therefore it can  be as significant    as 
 Eq.~(\ref{ebndy}). This term is however absent in our case due to its relation with the anomalies of the theory; it vanishes because $SU(2)$ is an anomaly-free group. Even in models where the CS term  can be present, like for example in $U(2)$ gauge theories, its effect could be as small as that of dimension 6-operators 
\cite{Hata:2007mb}, even though the situation is not completely clear. We leave the analysis of the CS effects for the future.

The issue of calculability is   
particularly compelling in our case, 
and it is different from  what happens
in  the 4D Skyrme model in which all operators give comparable contributions. 
\footnote{When this is the case, not only the effective theory does not give us any  quantitative information on the soliton, but also its very existence is doubtful.}  
The problem, in the case of the  4D skyrmions, comes from the fact that a higher-dimensional operator of the form $\alpha\left(\partial U\right)^4$ must be added to the Goldstone kinetic term $F_\pi^2\left(\partial U\right)^2$ in order to obtain a regular solution of the EOM \cite{Skyrme:1961vq, Adkins:1983ya}. The value of $\alpha$ estimated by NDA is $F_\pi^2/\Lambda^2$, where 
$\Lambda\sim 4 \pi F_\pi$ is the cut-off of the chiral lagrangian. 
We then see that  $F_\pi^2$ factorizes in front of the action and the cut-off $\Lambda$ is the only dimensionful quantity which enters into the EOM, 
setting  the radius of the skyrmion  to  ${\rho_0}\sim 1/\Lambda$. 
In our 5D case 
also a higher-dimensional operator is needed to stabilize the soliton. Its
 coefficient is given by $\alpha\sim M_5/\Lambda_5$ and then also 
 the 5D coupling $M_5$  factorizes in front of the action.
Nevertheless,
 our soliton configuration not only depends  on the 5D cut-off $\Lambda_5$, but also on the conformal lenght   $L$.
 This is the crucial difference from  the 4D Skyrme model;
 a combination of $L$ and $\Lambda_5$   sets the size of the 5D soliton 
 fixing it to a scale larger than $1/\Lambda_5$. 

\section{Conclusions}

We have studied    skyrmion configurations 
arising from compact five dimensional models.
We have shown that the size of these  skyrmions 
is  stabilized  by  the presence of IR-boundary kinetic
terms.
This size  is always larger than the inverse of the 5D cut-off scale $1/\Lambda_5$,
and therefore   consistent  within our 5D effective theory.
This is different from   previous 5D models \cite{Hong:2007kx,Hata:2007mb}
  in which the skyrmion size was found to be of  order  $1/\Lambda_5$.
We have  numerically obtained the skyrmion configurations
for different values  of $\alpha$ (the coefficient of the IR-boundary kinetic term),
and  calculated their size and energy.
Although the size of the skyrmions     depends strongly on $\alpha$,  
their energy is quite model-independent.
By varying $\alpha$ 
these skyrmion configurations  smoothly  interpolates  
 between  small 4D  instantons and  
 4D skyrmions made of Goldstones and a massive gauge boson.

The existence of    stable skyrmion configurations 
 in extra dimensional models rises    different  
phenonemological issues. 
For example,   theories of electroweak symmetry breaking arising from 
extra dimensions  will  have this type of stable configurations, and therefore 
the  analysis of their cosmological consequences are of great importance.  
For 5D Higgsless \cite{Csaki:2003zu}
or composite Higgs  models   \cite{Contino:2006qr}
in which 
$F_\pi= 246/\epsilon$  GeV (where $\epsilon\leq 1$) and $m_\rho\simeq  1.2/\epsilon$ TeV,
we find a soliton energy  $M_N\sim (2.5-3)/\epsilon$ TeV.  These TeV stable particles
are  possible   dark  matter candidates,  therefore the
precise determination of their  relic abundances is of important  phenomenological interest. 
These configurations can also be useful    
in holographic models of 4D strong  interactions.
The AdS/CFT correspondence tells us that these 5D solitons  correspond  to   the "baryons"
of the dual theory. 
The  5D AdS model studied here has  already been  proposed as 
a holographic model of QCD, giving  predictions for the  meson spectrum and couplings
in good agreement with the experimental data \cite{Erlich:2005qh}. 
The skyrmion found here gives an approximate value for the  mass of  the proton. We find
  $M_N\simeq 500-650$ GeV, too low compared with the experimental value
$\sim 1$ GeV.
We must notice, however, that in our analysis
we have not included  the 5D CS term, responsible in holographic QCD
models of  the Wess-Zumino-Witten (WZW) term.
It is known that the presence of the WZW  term has  important  
impact on the skyrmion mass
\cite{Adkins:1983nw}, so we can expect that  including the CS 
in our analysis can enhance
our  prediction of the proton mass.
These and other phenomenological questions  deserve a further
analysis that we leave for a  future work.

{\bf
 Acknowledgments} 

\noindent We would like to thank 
Gia Dvali and   Jose Antonio Carrillo
 for valuable conversations.
 We also thank   Josep Maria  Mondelo for helping us with FEMLAB.   
This work has been partly supported by
the FEDER  Research Project FPA2005-02211,
 the DURSI Research Project SGR2005-00916 and 
the European Union
under contract MRTN-CT-2004-503369 and
MRTN-CT-2006-035863.


\section*{Appendix: The instanton energy in compact and  warped spaces}

In the 
2D notation used in this paper the BPST instanton with center at $({\bf x}=0,z=z_\IR)$ and size $\rho$  corresponds to  \cite{Manton:1978fr}
\beq
\begin{array}{ll}
\phi_1\,=\,r\,\displaystyle\frac{\partial_z\Phi}\Phi\,\equiv\,\bar\phi_1(\rho)\,,& \phi_2\,=\,-1-r\,\displaystyle\frac{\partial_r\Phi}\Phi\,\equiv\,\bar\phi_2(\rho)\,,\\
A_1\,=\,\displaystyle\frac{\partial_z\Phi}\Phi\,\equiv\,\bar A_1(\rho)\,,&
A_2\,=\,-\,\displaystyle\frac{\partial_r\Phi}\Phi\,\equiv\,\bar A_2(\rho)\,,
\end{array}
\label{inst}
\eeq
where
\beq
\Phi\,=\,\frac{1}{\rho^2+r^2+(z-z_\IR)^2}\, .
\eeq
The singular instanton of size $\rho\rightarrow0$ solves, up to a gauge transformation, the variational problem defined by the energy Eq.~(\ref{en2d}) and the 
boundary conditions Eqs.~(\ref{boundary condition}) and (\ref{boundary conditionr0}). 
Indeed, it has the same topological charge ($Q=1$) of the solution we were looking for and, since its 2D energy density is a $\delta$-function 
at $r=0$, $z=z_\IR$, it has the minimal energy $E=8\pi^2M_5$ which saturates the lower bound of Eq.~(\ref{bound}).

We can explicitly check that the instanton Eq.~(\ref{inst}) fullfills,  for any $\rho$,  the boundary conditions at $r=0$ and $z=z_\IR$
given in Eqs.~(\ref{boundary condition}) and (\ref{boundary conditionr0}). 
At the other two boundaries the instanton of size $\rho\rightarrow0$
has different boundary conditions: 
\beq
z=z_\UV\,:\,\left\{\begin{array}{l}\phi=\bar\phi(0)=\displaystyle{-ie^{i\,\beta}}
\\A_1=\bar A_1(0)=\partial_{1}\beta\\
\partial_{\bar\mu} A^{\bar\mu}=0\end{array}\right.\, ,\;\;\;\;\;\;
r\rightarrow\infty\,:\,\left\{\begin{array}{l}
\phi=\bar\phi(0)=\displaystyle{-ie^{i\,\beta}}
\\A_2=\bar A_2(0)=\partial_{2}\beta\\
\partial_{\bar\mu} A^{\bar\mu}=0\end{array}\right.\,,
\label{newboundary condition}
\eeq
where $\beta\,=\,2\arctan\left[r/(z_\IR-z)\right]$. These  boundary conditions are, as anticipated, topologically equivalent to those in Eq.~(\ref{boundary condition}) and then one can convert one into the other by a gauge transformation. Instead of gauge rotating the instanton of $\rho=0$  to make it fulfill Eq.~(\ref{boundary condition}),  it is more convenient to rephrase our original problem in the new gauge
and the new boundary conditions   Eq.~(\ref{newboundary condition}).

Let us consider instantons of small but non-vanishing size.
We  would like to compute the energy $E(\rho)$ of such configurations. 
We already know that $E(\rho)$ will have an absolute minimum at $\rho=0$; this means that a small instanton  is subject to an ''attractive force'' which tends to shrink its size to zero.
 By computing $\partial_\rho E(\rho)$ one can measure the strength  of this force. It is important to remark that there are two different effects which make the force arise. The first one, which is essentially local, is due to the curvature of the space and therefore it is only present   in warped   spaces. 
It pushes the instanton 
to be localized as much as possible at $z_\IR$ where  
the    warp factor has its minimal value and then it is  energetically favorable.
The second effect is non-local and is due to the presence of the boundary at $z_\UV$ which confines the solution into a finite volume. At the  more technical level the two effects come, respectively, from the fact that the instanton fails to fulfill  the bulk EOM when
the space is  warped
and the boundary condition Eq.~(\ref{newboundary condition}) at $z_\UV$.
It cannot then be a  stable configuration.

To compute $E(\rho)$ let us first try to substitute the instanton configuration in the energy functional Eq.~(\ref{en2d}). 
We call $E_B$ the ''bulk'' contribution to the energy which we obtain in this way. 
We will see later that this  is not the only contribution to $E(\rho)$.
For a generic warp factor we have
 \footnote{We are considering, everywhere in this Appendix, $z_\UV\neq0$ because for $z_\UV=0$ the energy of the instanton in the AdS slice diverges. Our numerical computations show however that the true solution is widely insensitive to the position of the UV-boundary  and the results we derive here can be applyied for $z_\UV=0$ as well.
 }
\beq
E_B(\rho)\,=\,12\,\pi^2 M_5\int_{-L/\rho}^{0}dy\,\frac{a(z_\IR+\rho y)}{(1+y^2)^{5/2}}\,=\,12\,\pi^2 M_5\sum_n \rho^nc_n(\rho/L)\frac{a^{(n)}(z_\IR)}{n!}\, ,
\label{eB}
\eeq
where we have expanded $a(z)$ in Taylor series around $z=z_\IR$ and 
\beq
c_n(\rho/L)=\int_{-L/\rho}^{0}dy\, \frac{y^n}{(1+y^2)^{5/2}}\, .
\eeq
The leading term of $E_B$ in the small $\rho$ expantion   comes from the $n=0$ term which reads
\beq
12\,\pi^2 M_5c_0(\rho/L)\simeq 8\,\pi^2 M_5\left(1\,-\,\frac38\frac{\rho^4}{L^4}+\mathcal{O}(\rho^6/L^6)\right),
\label{n0}
\eeq
where we have also expanded $c_0(\rho/L)$ around $\rho=0$ 
and have kept, for future convenience, the first correction  of order $\rho^4$. 
A linear contribution to the energy can only come from the $n=1$ term and is given by
\beq
-4\,\pi^2 M_5\rho\,a^{(1)}(z_\IR)\,\left(1\,+\,\mathcal{O}(\rho^3/L^3)\right)\,.
\eeq
If $a^{(1)}$ is non zero, this term  must be negative since the warp 
factor is a decreasing function of $z$.
Therefore  the above term  gives, as expected, an attractive force which makes the instanton shrink. In the case of  AdS, Eq.~(\ref{adsmetric}), we have  $a^{(1)}=-1/L$ 
and we obtain the result of Eq.~(\ref{adsen}). 
For a space with $a^{(1)}=0$ the linear contribution vanishes and the leading correction to the energy, which comes from the $n=2$ term, is of $\mathcal{O}(\rho^2/L^2)$. 
Making the space less and less warped near $z_\IR$ we are increasing  
the power of $\rho/L$ of the instanton energy, 
 and therefore making the attractive force weaker and weaker at small $\rho$. 
 This process however stops at order $\rho^4/L^4$.
 The $n= 4$ term is proportional to $c_4\sim \ln\rho$,
 and for $n> 4$    we have  $c_n\sim 1/\rho^{n-4}$,
  so   all terms $n\geq 4$ contribute at the order $\rho^4/L^4$ to the energy. 
Therefore we   expect that   the weakest possible force
will be obtained from  a energy 
of $\mathcal{O}(\rho^4/L^4)$ that  is, as we will now show, 
what actually happens in the case of flat space.

For flat space only the $n=0$ term appears in Eq.~(\ref{eB}) and $E_B$ is
given by Eq.~(\ref{n0}). We then immediatly see that $E_B$ cannot be the
total instanton energy. If so, the point $\rho=0$ would not be a minimum
but a maximum, and we would not find an attractive but a repulsive force.
The reason why there is  an extra contribution to the energy   is that the
instanton of finite size does not respect the boundary condition at $z_\UV$. Therefore, it
does not belong to the set of allowed field configurations on which our
variational problem is defined. In order to find the instanton energy we
need a different formulation of the variational problem in which the
allowed field configurations do not necessarily respect the boundary condition. The latter
will arise, like the EOM, from the minimization of a new energy functional
which is now defined to act on a more general set of configurations, to
which the instanton belongs. 
This new functional is given by Eq.~(\ref{en2d}) and the UV-boundary term
\footnote{We can obtain this result by using  Lagrange Multipliers  
that allows to include  boundary constraints in the functional to minimize.}
\be
E_\UV\,=\,8\pi M_5a(z_\UV)\int_0^\infty dr\,\left[\left(\phi-\bar\phi(0)\right)^* D_z\phi+\left(\phi-\bar\phi(0)\right) D_z\phi^*-\frac{r^2}{2}\left(A_1-\bar A_1(0)\right)F_{12}\right]_{z_\UV}\,.
\label{euv}
\ee
It is easily understood why the addition of $E_\UV$ makes the UV boundary conditions arise as EOM.
The  variation of Eq.~(\ref{en2d}) and (\ref{euv}) gives,
in addition to the usual bulk terms whose cancellation will give rise to the EOM, 
localized UV-terms of the form
 $\left(\phi-\bar\phi\right)^* \delta(D_z\phi)$ and $\left(A_1-\bar A_1\right)\delta(F_{12})$.
Requiring those terms to vanish enforces the boundary conditions in Eq.~(\ref{newboundary condition}).


We can finally compute the energy $E(\rho)$ of the finite size instanton. It is given by the sum of $E_B$ in Eq.~(\ref{eB}) and $E_\UV$ in Eq.~(\ref{euv}), in which of course we have to substitute the instanton configuration. The latter term reads
\be
E_\UV(\rho)\,=\,5\pi^2M_5a(z_\UV)\frac{\rho^4}{L^4}\,,
\ee
and gives, for a generic warped metric, a $\rho^4$ contribution to the energy. This contribution does not affect the result in the case of the AdS slice, since it is subleading, 
but it is crucial in flat space as it corrects the negative sign which we found in Eq.~(\ref{n0}). 
Adding  $E_B$ and $E_\UV$ we obtain Eq.~(\ref{flaten}).



\begin{thebibliography}{99}


\bibitem{Skyrme:1961vq}
  T.~H.~R.~Skyrme,
  Proc.\ Roy.\ Soc.\ Lond.\  A {\bf 260} (1961) 127.

\bibitem{Son:2003et}
  D.~T.~Son and M.~A.~Stephanov,
  Phys.\ Rev.\  D {\bf 69} (2004) 065020.

\bibitem{Hong:2007kx}
  D.~K.~Hong, M.~Rho, H.~U.~Yee and P.~Yi,
  Phys.\ Rev.\  D {\bf 76} (2007) 061901.

\bibitem{Hata:2007mb}
  H.~Hata, T.~Sakai, S.~Sugimoto and S.~Yamato,
  arXiv:hep-th/0701280.

\bibitem{Hong:2006ta}
  D.~K.~Hong, T.~Inami and H.~U.~Yee,
  Phys.\ Lett.\  B {\bf 646} (2007) 165;
  K.~Nawa, H.~Suganuma and T.~Kojo,
  Phys.\ Rev.\  D {\bf 75} (2007) 086003.

\bibitem{Hill:2001bt}
   C.~T.~Hill,
   Phys.\ Rev.\ Lett.\  {\bf 88} (2002) 04160.


\bibitem{Csaki:2003zu}
  C.~Csaki, C.~Grojean, L.~Pilo and J.~Terning,
  Phys.\ Rev.\ Lett.\  {\bf 92}, 101802 (2004);
  Y.~Nomura,
  JHEP {\bf 0311} (2003) 050;
  R.~Barbieri, A.~Pomarol and R.~Rattazzi,
  Phys.\ Lett.\  B {\bf 591}, 141 (2004).



\bibitem{Erlich:2005qh}
  J.~Erlich, E.~Katz, D.~T.~Son and M.~A.~Stephanov,
  Phys.\ Rev.\ Lett.\  {\bf 95} (2005) 261602;
  L.~Da Rold and A.~Pomarol,
  Nucl.\ Phys.\  B {\bf 721} (2005) 79.


\bibitem{Panico:2007qd}
  G.~Panico and A.~Wulzer,
  JHEP {\bf 0705} (2007) 060.

\bibitem{Chu:1996fr}
  We are following the notations of 
  C.~S.~Chu, P.~M.~Ho and B.~Zumino,
  Nucl.\ Phys.\  B {\bf 475}, 484 (1996)
  [arXiv:hep-th/9602093].

\bibitem{Witten:1976ck}
  E.~Witten,
  Phys.\ Rev.\ Lett.\  {\bf 38} (1977) 121.



\bibitem{comsol}
See http://www.comsol.com.






\bibitem{Igarashi:1985et}
    Y.~Igarashi, M.~Johmura, A.~Kobayashi, H.~Otsu, T.~Sato and S.~Sawada,
  Nucl.\ Phys.\  B {\bf 259} (1985) 721.


\bibitem{Adkins:1983ya}
  G.~S.~Adkins, C.~R.~Nappi and E.~Witten,
  Nucl.\ Phys.\  B {\bf 228} (1983) 552.


\bibitem{Contino:2006qr}
  R.~Contino, L.~Da Rold and A.~Pomarol,
  Phys.\ Rev.\  D {\bf 75} (2007) 055014.
  
\bibitem{Adkins:1983nw}
  G.~S.~Adkins and C.~R.~Nappi,
  Phys.\ Lett.\  B {\bf 137} (1984) 251.
  
\bibitem{Manton:1978fr}
  N.~S.~Manton,
  Phys.\ Lett.\  B {\bf 76} (1978) 111.

      
  
\end{thebibliography}
\end{document}